\documentclass[11pt]{article}%
\usepackage{amsfonts}
\usepackage{amsmath}
\usepackage{amssymb}
\usepackage{indentfirst}
\usepackage{graphicx}%
\providecommand{\U}[1]{\protect\rule{.1in}{.1in}}
\begin{document}

\title{Lovelock gravities from Born-Infeld gravity theory}
\author{P. K. Concha$^{1,2}$\thanks{patillusion@gmail.com},\thinspace\ N. Merino$^{3}%
$\thanks{nemerino@gmail.com},\thinspace\ E. K. Rodr\'{\i}guez$^{1,2}%
$\thanks{everodriguezd@gmail.com},\\{\small $^{1}$\textit{Departamento de Ciencias, Facultad de Artes Liberales y
}}\\{\small \textit{Facultad de Ingenier\'{\i}a y Ciencias, Universidad Adolfo
Ib\'{a}\~{n}ez,}}\\{\small Av. Padre Hurtado 750, Vi\~{n}a del Mar, Chile}\\{\small $^{2}$\textit{Instituto de Ciencias F\'{\i}sicas y Matem\'{a}ticas,
Universidad Austral de Chile,}}\\{\small Casilla 567, Valdivia, Chile}\\{\small $^{3}$\textit{Instituto de F\'{\i}sica, Pontificia Universidad
Cat\'{o}lica de Valpara\'{\i}so,} }\\{\small Casilla 4059, Valpara\'{\i}so, Chile}}
\maketitle

\begin{abstract}
We present a Born-Infeld gravity theory based on generalizations of Maxwell
symmetries denoted as $\mathfrak{C}_{m}$. We analyze different configuration
limits allowing to recover diverse Lovelock gravity actions in six dimensions.
Further, the generalization to higher even dimensions is also considered.

\end{abstract}

\vspace{-11cm}

\begin{flushright}
{\footnotesize UAI-PHY-16/09}
\end{flushright}

\vspace{10cm}

\section{Introduction}

It is a common assumption in theoretical physics that the spacetime may have
more than four dimensions.\ This requires a generalization of General
Relativity (GR) theory of gravity that includes general covariance and second
order field equations for the metric. Although the Einstein-Hilbert (EH)
action can be generalized to higher dimensions, the most general metric theory
of gravity, satisfying the criteria of general covariance and giving second
order field equations, is given by the Lanczos-Lovelock theory (LL)
\cite{Lovelock:1971yv,Lanczos}. The LL action is constructed as a polynomial
of degree $\left[  D/2\right]  $ in the Riemann curvature tensor
$R_{\ \beta\mu\nu}^{\alpha}\,$,%
\begin{equation}
I_{LL}[g]=\int d^{D}x\sum\limits_{k=0}^{\left[  D/2\right]  }\alpha
_{k}\mathcal{L}_{k}\,. \label{second_order1}%
\end{equation}
with
\begin{equation}
\mathcal{L}_{k}=\frac{1}{2^{k}}\,\sqrt{-g}\ \delta_{\nu_{1}...\nu_{2k}}%
^{\mu_{1}...\mu_{2k}}\ R_{\mu_{1}\mu_{2}}^{\nu_{1}\nu_{2}}\cdots R_{\mu
_{2k-1}\mu_{2k}}^{\nu_{2k-1}\nu_{2k}}\,, \label{second_order2}%
\end{equation}
and where $\alpha_{k}$ are arbitrary constants and $\delta_{\nu_{1}...\nu
_{2k}}^{\mu_{1}...\mu_{2k}}$ is the generalized Kronecker delta. Although the
EH action is contained in the LL action, the action with higher powers of the
curvature in $D>4$ are dynamically different from GR and are not
perturbatively related.

Using first order formulation, where the affin connection $\Gamma_{\mu\nu
}^{\lambda}$ is suppose to be independent from the metric $g_{\mu\nu}$, the LL
theory acquires, in general, torsional degrees of freedom. This can be easily
seen using Riemann-Cartan formulation of gravity in terms of the vielbein and
spin connection one-forms $\left(  e^{a}{\small \ ,}\omega^{ab}\right)  $,
where $a,b=0,1,\ldots,D$ are the local Lorentz indices. In that case the LL
action can be regarded as the most general $D$-form invariant under local
Lorentz transformations, constructed out of the vielbein $e^{a}$, the spin
connection $\omega^{ab}$ and their exterior derivatives without using the
Hodge dual \cite{Zumino,TeitelboimZ},%
\begin{equation}
I_{LL}[e,\omega]=\int\sum_{k=0}^{\left[  D/2\right]  }\alpha_{k}%
\mathcal{L}^{\left(  k\right)  }\,, \label{LL_RC}%
\end{equation}
where $\alpha_{k}$ are arbitrary constants which are not fixed from first
principles and%
\begin{equation}
\mathcal{L}^{\left(  k\right)  }=\epsilon_{a_{1}\cdots a_{D}}R^{a_{1}a_{2}%
}\cdots R^{a_{2k-1}a_{2k}}e^{a_{2k+1}}\cdots e^{a_{D}}\,. \label{LL_RC_2}%
\end{equation}
and where the Riemann curvature and torsion $2$-forms are defined as
$R^{ab}=d\omega^{ab}+\omega_{\ c}^{a}\omega^{cb}$ and $T^{a}=de^{a}%
+\omega_{\ b}^{a}e^{b}\,$, respectively. Then it is direct to note that a
torsional dynamical field equation will arise for $k\geq2$.

It is worth to notice that the relation between the Riemann-Cartan action
(\ref{LL_RC}) with the tensorial first order formalism, this latter formulated
in terms of the metric and affin connection $\left(  g_{\mu\nu},\Gamma_{\mu
\nu}^{\lambda}\right)  $, is given through $g_{\mu\nu}=\eta_{ab}e_{\mu}%
^{a}e_{\nu}^{b}$ and $\Gamma_{\mu\nu}^{\lambda}=\omega_{\nu}^{ab}%
e_{a}^{\lambda}e_{a\mu}+e_{a}^{\lambda}\partial_{\nu}e_{\mu}^{a}$. The last
expression which is related with the metricity condition $\nabla^{\Gamma
}g_{\mu\nu}=0,$ assures that the Riemann curvature and torsion expressed as
$2$-forms and tensors are essentially the same objects in both languages,
i.e.,%
\begin{equation}
R_{\ \ \mu\nu}^{ab}=e_{\alpha}^{a}e_{\beta}^{b}R_{\ \ \mu\nu}^{\alpha\beta
}\,,\ \ T_{\mu\nu}^{a}=e_{\alpha}^{a}T_{\mu\nu}^{\alpha}\,.
\label{curvatures_metricity}%
\end{equation}

In the present work we will deal with the language of differential forms,
since it makes calculation much more compact and because it is more suitable
to describe the gauge structure that the theory possesses.

As shown in Ref.~\cite{TZ}, requiring the theory to have the maximum possible
number of degrees of freedom fixes, in the first order formalism, the\textbf{
}$\alpha_{k}$\textbf{ }constants in (\ref{LL_RC}). In odd dimensions, the
Lagrangian becomes a Chern-Simons (CS) form \cite{Cham1,Cham2}, which is a
functional $\mathcal{L}_{CS}\left[  A\right]  $ of a gauge connection one-form
$A$ containing the vielbein and spin connection. The corresponding CS action
is invariant, up to a boundary term, under a bigger symmetry (dS, AdS or
Poincar\'{e} groups). In even dimensions the same requirement leads to a
Born-Infeld (BI) action which is also constructed in terms of the curvature
associated with the gauge connection, but it is locally invariant only under
the Lorentz subgroup.

Another interesting family, called Pure Lovelock (PL) gravity, has been
recently proposed in \cite{Cai:2006pq,Dadhich:2012ma,Dadhich:2012eg} as
another way of fixing the $\alpha$'s. It consists of only two terms, the
cosmological one and a single $p$-power in the curvature, with $p=1,\ldots
,N=\left[  \left(  D-1\right)  /2\right]  $. Remarkably, their black holes
solutions behave asymptotically like the ones in GR and like the dimensionally
continued black holes \cite{Banados:1993ur,Crisostomo:2000bb} near the horizon.

In Ref.~\cite{TeitelboimZ} it was suggested that the metric LL theory should
have the same degrees of freedom (dof) as the higher-dimensional EH gravity,
i.e., $D\left(  D-3\right)  /2$. However, the non-linearity of the theory
makes the symplectic matrix to change the rank with the backgrounds
\cite{Joel} generating extra local symmetries and decreasing degrees of
freedom in some of them. This behavior, typical for LL theories, has also been
found in Lovelock-Chern-Simons gravities \cite{BGH1,BGH2} and recently in PL
gravity \cite{Dadhich:2015ivt} where the number of dof changes with the
backgrounds between $0$ and $D\left(  D-3\right)  /2$. Besides, this property
is not only intrinsic of the Riemann sector, but also happens when torsional
degrees of freedom are considered. This is the case when one looks for charged
black hole solutions in CS supergravity \cite{GMMZ}.

On the other hand, the supersymmetric version of the LL theory is not known in
general, except for few cases such as the EH and CS ones. The existence of a
supersymmetric version in $D=5$ for non vanishing constants $\alpha_{1}$ and
$\alpha_{2}$ is discussed in Ref.~\cite{Deser:2011zk}, even though its
explicit form is still unknown. It has also been suggested in \cite{CDIMR}
that a supersymmetric version of PL theory might be constructed using new
symmetries obtained through expansion methods of Lie algebras
\cite{AIPV,Sexp,AMNT}. Indeed, those methods have already been used to
relate\textbf{ }diverse gravity theories. For example, it has been found that
even and odd-dimensional GR can be obtained as a special limit of BI and CS
theories, constructed with expansions of the $\mathfrak{so}\left(
D-1,2\right)  $ algebra \cite{GRCS,CPRS1,CPRS2,CPRS3,Gonzalez:2014tta}.

Recently, in Ref.~\cite{CDIMR}, it has been shown that the PL action in odd
dimensions can be obtained as a limit from a CS action based on a special
expansion of the $\mathfrak{so}\left(  D-1,2\right)  $ algebra, denoted by
$\mathfrak{C}_{m}$. Those symmetries were introduced in
Refs.c\cite{Sorokas,Sorokas2,DKGS,DFIMRSV,SS,CDMR,REMI} and can be regarded as
generalizations of the so called Maxwell algebra \cite{Schrader,BCR}, which
describes the symmetries of quantum fields in Minkowski space with the
presence of a constant electromagnetic field. Thus, for completeness and also
due to the growing interest in the effect of higher-curvature terms in the
holographic context (see for example
\cite{Brigante:2008gz,Aranguiz:2013cwa,Camanho:2013pda,Aranguiz:2015voa}), in
this work we will show that different Lovelock gravity actions in even
dimensions can be obtained from a BI action based on the $\mathfrak{C}_{m}$
algebra. We shall start by considering the six-dimensional spacetime since it
allows us to obtain a bigger variety of gravity theories. Indeed, by applying
in four dimensions the prescription presented here would lead only to the
Einstein gravity with cosmological constant term.

The present work is organized as follows: in Section 2 we briefly review the
BI gravity theory. Section 3 and 4 contain our main results. We present the
explicit expression for the six-dimensional BI type gravity action based on
the\textbf{ }$\mathfrak{C}_{7}$\textbf{ }two-form curvature. The general setup
in order to derive different Lovelock gravity action in a particular limit is
given. We conclude our work by providing the generalization to higher even dimensions.

\section{Brief review about Born-Infeld gravity theory}

As was previously pointed out the Lanczos-Lovelock theory refers to a family
parametrized by a set of real coefficients $\alpha_{k}$, which are not fixed
from first principles. To require the theory possess the largest possible
number of degrees of freedom, fixes the $\alpha_{k}$ parameters in terms of
the gravitational and the cosmological constants \cite{TZ}.\ As a result, in
even dimensions the action has a Born-Infeld form invariant only under local
Lorentz rotations, in the same way as the EH action. In this section, we
review the main aspects of the BI gravity theory. As was shown in
Ref.~\cite{TZ}, choosing the coefficients as
\begin{equation}
\alpha_{k}=\alpha_{0}\left(  2\gamma\right)  ^{k}\left(
\begin{array}
[c]{c}%
n\\
k
\end{array}
\right)  \,,\label{alphabi}%
\end{equation}
with $0\leq k\leq n$ and
\[
\alpha_{0}=\frac{\kappa}{\left(  2n\right)  l^{2n}}\,,\ \ \ \ \gamma
=-sign\left(  \Lambda\right)  \frac{l^{2}}{2}\,,
\]
the LL Lagrangian leads to the so-called Lovelock-Born-Infeld (LBI)
Lagrangian\cite{DG, DJT} in $D=2n$%
\begin{equation}
\mathcal{L}_{BI}^{2n}=\frac{\kappa}{2n}\epsilon_{a_{1}\cdots a_{2n}}\bar
{R}^{a_{1}a_{2}}\cdots\bar{R}^{a_{2n-1}a_{2n}}\,,\label{LBI}%
\end{equation}
where $\bar{R}^{ab}=R^{ab}+\frac{1}{l^{2}}e^{a}e^{b}$ corresponds to the $AdS$
curvature. Here $R^{ab}$ is the usual Lorentz curvature and $l$ is a length
parameter. Let us notice that the Lagrangian (\ref{LBI}) is the Pfaffian of
the $2$-form $\bar{R}^{ab}$ and can be rewritten as,%
\begin{equation}
\mathcal{L}_{BI}^{2n}=2^{n-1}\left(  n-1\right)  !\kappa\sqrt{\det\left(
R^{ab}+\frac{1}{l^{2}}e^{a}e^{b}\right)  }\,.\label{LL16}%
\end{equation}
which remind us the Born-Infeld electrodynamics Lagrangian.

The BI gravity Lagrangian, which is basically contructed under the requirement
of having a unique maximally degenerate AdS vacuum, has the advantage of
having well defined black holes configurations. The family of gravity actions
constructed with the coefficients (\ref{alphabi}) up to a certain
fixed\textbf{ }$k$, with $0\leq k\leq n$, are characterized by the fact that
they have a unique $k$-order degenerate AdS vacuum, with the BI Lagrangian
case being described by\textbf{ }$k=n$\textbf{. }As shown in
\cite{Banados:1993ur, Crisostomo:2000bb}, such family (which includes EH and
EGB gravities for $k=1,2$) is free of degeneracies in the static sherically
symmetric sector of the space of solutions, i.e., they have well defined black
holes configurations. This not occurs for the Lovelock theory which, for
arbitrary $\alpha_{k}$ constants, the field equations do not determine
completely the components of the curvature and torsion\textbf{ }in the static
sector. Besides, the BI gravity Lagrangian possess a large number of appealing
features like cosmological models, black hole solutions, etc.

It is important to note that the Lagrangian (\ref{LBI}) is invariant only
under local Lorentz transformations and not under the $AdS$ group. In this
way, in $D=2n$ \ the Levi-Civita symbol $\epsilon_{a_{1}\cdots a_{2n}}$ in
(\ref{LBI}) can be regarded as the only invariant tensor under the Lorentz
group $SO(2n-1,1).$ This choice of the invariant tensor, which is necessary in
order to reproduce a non-trivial action principle, breaks the full AdS
symmetry to their Lorentz subgroup. In fact, the BI gravity action can be
written as follows%
\begin{equation}
I_{BI}^{2n}=\kappa\int\left\langle F\wedge\cdots\wedge F\right\rangle
=\kappa\int F^{A_{1}}\wedge\cdots\wedge F^{A_{n}}\left\langle T_{A_{1}}\cdots
T_{A_{n}}\right\rangle \,. \label{BIaction}%
\end{equation}
Here $F$ is the $AdS$ $2$-form curvature
\[
F=F^{A}T_{A}=\frac{1}{2}\left(  R^{ab}+\frac{1}{l^{2}}e^{a}e^{b}\right)
J_{ab}+\frac{1}{l}T^{a}P_{a}\,,
\]
and $\left\langle T_{A_{1}}\cdots T_{A_{n}}\right\rangle $ is chosen as an
invariant tensor for the Lorentz group only,
\begin{equation}
\left\langle T_{A_{1}}\cdots T_{A_{n}}\right\rangle =\left\langle
J_{a_{1}a_{2}}\cdots J_{a_{2n-1}a_{2n}}\right\rangle =\frac{2^{n-1}}%
{n}\epsilon_{a_{1}\cdots a_{2n}}\,.
\end{equation}
Then the action (\ref{BIaction}) can be expressed as
\begin{equation}
I_{BI}^{2n}=\int%
{\displaystyle\sum\limits_{k=0}^{n}}
\frac{\kappa}{2n}\left(
\begin{array}
[c]{c}%
n\\
k
\end{array}
\right)  l^{2k-2n}\epsilon_{a_{1}\cdots a_{2n}}R^{a_{1}a_{2}}\cdots
R^{a_{2k-1}a_{2k}}e^{a_{2k+1}}\cdots e^{a_{2n}}\,.
\end{equation}

In $D=4$ the BI action is written as a particular linear combination of the
standard Einstein-Hilbert action with cosmological constant and the Euler
density,
\begin{equation}
I_{BI}^{4D}=\frac{\kappa}{4}\int\epsilon_{abcd}\left(  R^{ab}R^{cd}+\frac
{2}{l^{2}}R^{ab}e^{c}e^{d}+\frac{1}{l^{4}}e^{a}e^{b}e^{c}e^{d}\right)  \,.
\label{LL17}%
\end{equation}
\ 

In the same way,\textbf{ }the Lovelock coefficients\textbf{ }$\alpha
_{0},\alpha_{1},\alpha_{2}$\textbf{ }and\textbf{ }$\alpha_{3}$\textbf{ }are
chosen so that the $D=6$ BI gravity action is given by
\begin{equation}
I_{BI}^{6D}=\frac{\kappa}{6}\int\epsilon_{abcdef}\left(  R^{ab}R^{cd}%
R^{ef}+\frac{3}{l^{2}}R^{ab}R^{cd}e^{e}e^{f}+\frac{3}{l^{4}}R^{ab}e^{c}%
e^{d}e^{e}e^{f}+\frac{1}{l^{6}}e^{a}e^{b}e^{c}e^{d}e^{e}e^{f}\right)  \,.
\end{equation}

\section{D=6 Lovelock gravity actions from Born-Infeld type theory}

In this section, we show the explicit construction of a BI type theory based
on enlarged symmetries and its relation to different six-dimensional Lovelock
gravity actions.

\subsection{Why \thinspace$\mathfrak{C}_{m}$ algebras?}

Our objective requires to find a symmetry which allows to separate each term
of the original Lovelock Lagrangian in different sectors. Thus, under a
specific limit, the unwanted sector can be avoided leading to interesting
gravity actions. To this purpose, every Lovelock term should be originated by
different components of an invariant tensor leading to a Born-Infeld type
action. These desired properties have origin in the $\mathfrak{so}\left(
D-1,2\right)  \oplus\mathfrak{so}\left(  D-1,1\right)  $ Lie algebra
\footnote{Also known as AdS-Lorentz algebra.}%
\thinspace\cite{Sorokas, Sorokas2, DKGS, DFIMRSV} which has been generalized,
using the abelian semigroup expansion method ($S$-expansion) \cite{Sexp}, to a
family of Maxwell type algebras denoted as $\mathfrak{C}_{m}\,$\cite{SS,CDMR}.
Their supersymmetric extensions have also been constructed in Refs.~\cite{CR1,
CRS}.

As was shown in Ref.~\cite{CDIMR} the $\mathfrak{C}_{m}$ algebras are obtained
from $AdS$ considering $S_{M}^{\left(  m-2\right)  }=\left\{  \lambda
_{0},\lambda_{1},\dots,\lambda_{m-2}\right\}  $ as the relevant semigroup,
whose multiplication law is given by%
\begin{equation}
\lambda_{\alpha}\lambda_{\beta}=\left\{
\begin{array}
[c]{c}%
\lambda_{\alpha+\beta},\text{ \ \ \ \ \ \ \ \ \ \ if }\alpha+\beta\leq
m-2\,,\\
\lambda_{\alpha+\beta-2\left[  \frac{m-1}{2}\right]  }\,,\text{ \ \ if }%
\alpha+\beta>m-2\,.
\end{array}
\right.  \label{ML}%
\end{equation}
After extracting a resonant subalgebra, one finds the $\mathfrak{C}_{m}$
algebra whose generators satisfy the following commutation relations%
\begin{align}
\left[  J_{ab,\left(  i\right)  },J_{cd,\left(  j\right)  }\right]   &
=\eta_{bc}J_{ad,\left(  i+j\right)  \operatorname{mod}\left[  \frac{m-1}%
{2}\right]  }-\eta_{ac}J_{bd,\left(  i+j\right)  \operatorname{mod}\left[
\frac{m-1}{2}\right]  }\nonumber\\
&  -\eta_{bd}J_{ac,\left(  i+j\right)  \operatorname{mod}\left[  \frac{m-1}%
{2}\right]  }+\eta_{ad}J_{bc,\left(  i+j\right)  \operatorname{mod}\left[
\frac{m-1}{2}\right]  }\,,\nonumber\\
\left[  J_{ab,\left(  i\right)  },P_{a,\left(  k\right)  }\right]   &
=\eta_{bc}P_{a,\left(  i+k\right)  \operatorname{mod}\left[  \frac{m-1}%
{2}\right]  }-\eta_{ac}P_{b,\left(  i+k\right)  \operatorname{mod}\left[
\frac{m-1}{2}\right]  }\text{\thinspace},\nonumber\\
\left[  P_{a,\left(  i\right)  },P_{b,\left(  k\right)  }\right]   &
=J_{ab,\left(  i+k+1\right)  \operatorname{mod}\left[  \frac{m-1}{2}\right]
}\,.
\end{align}
The new generators $\left\{  J_{ab,\left(  i\right)  },P_{a,\left(  k\right)
}\right\}  $ are related to the $\mathfrak{so}\left(  5,2\right)  $ ones
$\left\{  \tilde{J}_{ab},\tilde{P}_{a}\right\}  $ through%
\begin{align*}
J_{ab,\left(  i\right)  }  &  =\lambda_{2i}\otimes\tilde{J}_{ab}\,,\\
P_{a,\left(  k\right)  }  &  =\lambda_{2k+1}\otimes\tilde{P}_{a}\,,
\end{align*}
with $i=0,1,\dots,\left[  \frac{m-2}{2}\right]  $ and $k=0,1,\dots,\left[
\frac{m-3}{2}\right]  $.

The two-form curvature $F=dA+A\wedge A$ for the $\mathfrak{C}_{m}$ algebra is
given by%
\begin{equation}
F=F^{A}T_{A}=\frac{1}{2}\sum_{i}\mathcal{R}^{ab,\left(  i\right)
}J_{ab,\left(  i\right)  }+\frac{1}{\ell}\sum_{k}R^{a,\left(  k\right)
}P_{a,\left(  k\right)  }\,, \label{G2F}%
\end{equation}
where%
\begin{align*}
\mathcal{R}^{ab,\left(  i\right)  }  &  =d\omega^{ab,\left(  i\right)
}+\delta_{\left(  j+l\right)  \operatorname{mod}[\frac{m-1}{2}]}^{i}%
\omega_{\text{ }c}^{a,\left(  j\right)  }\omega^{cb,\left(  l\right)  }%
+\frac{1}{\ell^{2}}\delta_{\left(  p+q+1\right)  \operatorname{mod}\left[
\frac{m-1}{2}\right]  }^{i}e^{a,\left(  p\right)  }e^{b,\left(  q\right)
}\,,\\
R^{a,\left(  k\right)  }  &  =de^{a,\left(  k\right)  }+\delta_{\left(
j+p\right)  \operatorname{mod}[\frac{m-1}{2}]}^{k}\omega_{\text{ }%
c}^{a,\left(  j\right)  }e^{c,\left(  p\right)  }\,.
\end{align*}

Let us note that $\omega^{ab,\left(  0\right)  }$ and $e^{a,\left(  0\right)
}$ correspond to the spin connection $\omega^{ab}$ and the vielbein $e^{a}$, respectively.

In order to build a six-dimensional BI type gravity action based on the
$\mathfrak{C}_{m}$ two-form curvature, we require the explicit expression of
the invariant tensor. \ Fortunately, the $S$-expansion method offers the
possibility of deriving the invariant tensor for the expanded algebra from the
original one. \ Indeed, following the Theorem VII.2 of Ref.~\cite{Sexp}, the
non-vanishing components of an invariant tensor are given by%
\begin{equation}
\left\langle J_{ab,\left(  i\right)  }J_{cd,\left(  l\right)  }J_{ef,\left(
m\right)  }\right\rangle =\frac{4}{3}\sigma_{2j}\delta_{\left(  i+l+m\right)
\operatorname{mod}\left[  \frac{m-1}{2}\right]  }^{j}\epsilon_{abcdef}\,\,,
\label{ITcm}%
\end{equation}
where $\sigma_{2j}$ are arbitrary constants. As in the original BI case, this
choice of the invariant tensor breaks the $\mathfrak{C}_{m}$ group to a
Lorentz type subgroup generated by $\left\{  J_{ab,\left(  i\right)
}\right\}  $.

Since the original Lovelock terms will arise in the BI type action through
$\mathcal{R}^{ab,\left(  0\right)  }$ and $\mathcal{R}^{ab,\left(  1\right)
}$, it is straightforward to know from (\ref{ITcm}) in which sector every
Lovelock term will appear. The following table clarifies this point:%
\[%
\begin{tabular}
[c]{lcccc}
& $\sigma_{0}$ & $\sigma_{2}$ & $\sigma_{4}$ & $\sigma_{6}$\\\cline{2-5}%
$\mathfrak{C}_{4}$ & \multicolumn{1}{|c}{$RRR$} &
\multicolumn{1}{|c}{$RRee,Reeee,eeeeee$} & \multicolumn{1}{|c}{} &
\multicolumn{1}{|c|}{}\\\cline{2-5}%
$\mathfrak{C}_{5}$ & \multicolumn{1}{|c}{$RRR,Reeee$} &
\multicolumn{1}{|c}{$RRee,eeeeee$} & \multicolumn{1}{|c}{} &
\multicolumn{1}{|c|}{}\\\cline{2-5}%
$\mathfrak{C}_{6}$ & \multicolumn{1}{|c}{$RRR$} &
\multicolumn{1}{|c}{$RRee,eeeeee$} & \multicolumn{1}{|c}{$Reeee$} &
\multicolumn{1}{|c|}{}\\\cline{2-5}%
$\mathfrak{C}_{7}$ & \multicolumn{1}{|c}{$RRR,eeeeee$} &
\multicolumn{1}{|c}{$RRee$} & \multicolumn{1}{|c}{$Reeee$} &
\multicolumn{1}{|c|}{}\\\cline{2-5}%
$\mathfrak{C}_{m\geq8}$ & \multicolumn{1}{|c}{$RRR$} &
\multicolumn{1}{|c}{$RRee$} & \multicolumn{1}{|c}{$Reeee$} &
\multicolumn{1}{|c|}{$eeeeee$}\\\cline{2-5}%
\end{tabular}
\ \ \ \
\]
As $RRR=\epsilon_{abcdef}R^{ab}R^{cd}R^{ef}$ corresponds to a boundary term,
the minimal algebra which allows to separate each Lovelock term in different
sectors of the BI type action corresponds to the $\mathfrak{C}_{7}$ algebra.

\subsection{D=6 Lovelock gravity actions and $\mathfrak{C}_{7}$ algebra}

\qquad Considering the $\mathfrak{C}_{7}$ algebra, we present different limits
and conditions on the $\sigma$'s leading to various Lovelock gravity actions
in $D=6$.

Let us first consider the $\mathfrak{C}_{7}$-valuated connection one-form
\begin{equation}
A=\frac{1}{2}\omega^{ab}J_{ab}+\frac{1}{\ell}e^{a}P_{a}+\frac{1}{2}%
k^{ab}Z_{ab}+\frac{1}{\ell}h^{a}Z_{a}+\frac{1}{2}\tilde{k}^{ab}\tilde{Z}%
_{ab}+\frac{1}{\ell}\tilde{h}^{a}\tilde{Z}_{a}\,,
\end{equation}
and the associated curvature two-form $F=dA+A\wedge A$%
\begin{equation}
F=F^{A}T_{A}=\frac{1}{2}\mathcal{R}^{ab}J_{ab}+\frac{1}{\ell}R^{a}P_{a}%
+\frac{1}{2}F^{ab}Z_{ab}+\frac{1}{\ell}H^{a}Z_{a}+\frac{1}{2}\tilde{F}%
^{ab}\tilde{Z}_{ab}+\frac{1}{\ell}\tilde{H}^{a}\tilde{Z}_{a}\,, \label{curv}%
\end{equation}
where%
\begin{align*}
\mathcal{R}^{ab}  &  =R^{ab}+2k_{\text{ }c}^{\left[  a\right\vert }\tilde
{k}^{c\left\vert b\right]  }+\frac{1}{\ell^{2}}\left(  h^{a}h^{b}+2e^{a}%
\tilde{h}^{b}\right)  \,,\\
R^{a}  &  =T^{a}+k_{\text{ }c}^{a}\tilde{h}^{c}+\tilde{k}_{\text{ }c}^{a}%
h^{c}\,,\\
F^{ab}  &  =Dk^{ab}+\tilde{k}_{\text{ }c}^{a}\tilde{k}^{cb}+\frac{1}{\ell^{2}%
}\left(  e^{a}e^{b}+2h^{a}\tilde{h}^{b}\right)  \,,\\
H^{a}  &  =Dh^{a}+k_{\text{ }c}^{a}e^{c}+\tilde{k}_{\text{ }c}^{a}\tilde
{h}^{c}\,,\\
\tilde{F}^{ab}  &  =D\tilde{k}^{ab}+2k_{\text{ }c}^{a}k^{cb}+\frac{1}{\ell
^{2}}\left(  \tilde{h}^{a}\tilde{h}^{b}+2e^{a}h^{b}\right)  \,,\\
\tilde{H}^{a}  &  =D\tilde{h}^{a}+k_{\text{ }c}^{a}h^{c}+\tilde{k}_{\text{ }%
c}^{a}e^{c}\,.
\end{align*}
Here $D=d+\omega$ denotes the Lorentz covariant exterior derivative and
$R^{ab}$ is the usual Lorentz curvature.

Following eq.\thinspace(\ref{ITcm}), it is possible to show that the only
non-vanishing components of an invariant tensor required to build a BI gravity
action, are given by%
\begin{equation}%
\begin{tabular}
[c]{ll}%
$\left\langle J_{ab}J_{cd}J_{ef}\right\rangle =\frac{4}{3}\sigma_{0}%
\epsilon_{abcdef\,}$ & $\left\langle Z_{ab}Z_{cd}Z_{ef}\right\rangle =\frac
{4}{3}\sigma_{0}\epsilon_{abcdef}$\\
$\left\langle J_{ab}Z_{cd}\tilde{Z}_{ef}\right\rangle =\frac{4}{3}\sigma
_{0}\epsilon_{abcdef}$ & $\left\langle \tilde{Z}_{ab}\tilde{Z}_{cd}\tilde
{Z}_{ef}\right\rangle =\frac{4}{3}\sigma_{0}\epsilon_{abcdef}$\\
$\left\langle J_{ab}J_{cd}Z_{ef}\right\rangle =\frac{4}{3}\sigma_{2}%
\epsilon_{abcdef}$ & $\left\langle J_{ab}\tilde{Z}_{cd}\tilde{Z}%
_{ef}\right\rangle =\frac{4}{3}\sigma_{2}\epsilon_{abcdef}$\\
$\left\langle Z_{ab}Z_{cd}\tilde{Z}_{ef}\right\rangle =\frac{4}{3}\sigma
_{2}\epsilon_{abcdef}$ & $\left\langle Z_{ab}\tilde{Z}_{cd}\tilde{Z}%
_{ef}\right\rangle =\frac{4}{3}\sigma_{4}\epsilon_{abcdef}$\\
$\left\langle J_{ab}J_{cd}\tilde{Z}_{ef}\right\rangle =\frac{4}{3}\sigma
_{4}\epsilon_{abcdef}$ & $\left\langle J_{ab}Z_{cd}Z_{ef}\right\rangle
=\frac{4}{3}\sigma_{4}\epsilon_{abcdef}$%
\end{tabular}
\ \ \ \label{TIc7}%
\end{equation}
where $\sigma_{0}$, $\sigma_{2}$ and $\sigma_{4}$ are dimensionless arbitrary
constants. \ As was previously mentioned, this choice of invariant tensor
breaks the $\mathfrak{C}_{7}$ algebra to its Lorentz type subalgebra
$\mathfrak{L}^{\mathfrak{C}_{7}}$ generated by $\left\{  J_{ab},Z_{ab}%
,\tilde{Z}_{ab}\right\}  $.

Then, considering the curvature two-form (\ref{curv}) and the non-vanishing
components of the invariant tensor (\ref{TIc7}) in the general expression of a
six-dimensional Born-Infeld type gravity action%
\[
I_{\mathfrak{L}^{\mathfrak{C}_{7}}\mathfrak{-}BI}^{6D}=\kappa\int\left\langle
FFF\right\rangle =\kappa\int F^{A}F^{B}F^{C}\left\langle T_{A}T_{B}%
T_{C}\right\rangle \text{\thinspace\thinspace},
\]
we find%
\begin{align}
I_{\mathfrak{L}^{\mathfrak{C}_{7}}-BI}^{6D}=\frac{\kappa}{6}\epsilon
_{abcdef}\int &  \sigma_{0}\left[  \mathcal{R}^{ab}\mathcal{R}^{cd}%
\mathcal{R}^{ef}+6\mathcal{R}^{ab}F^{cd}\tilde{F}^{ef}+F^{ab}F^{cd}%
F^{ef}+\tilde{F}^{ab}\tilde{F}^{cd}\tilde{F}^{ef}\right] \nonumber\\
&  +\sigma_{2}\left[  3\mathcal{R}^{ab}\mathcal{R}^{cd}F^{ef}+3\mathcal{R}%
^{ab}\tilde{F}^{cd}\tilde{F}^{ef}+3F^{ab}F^{cd}\tilde{F}^{ef}\right]
\nonumber\\
&  +\sigma_{4}\left[  3\mathcal{R}^{ab}\mathcal{R}^{cd}\tilde{F}%
^{ef}+3\mathcal{R}^{ab}F^{cd}F^{ef}+3F^{ab}\tilde{F}^{cd}\tilde{F}%
^{ef}\right]  \,\,. \label{I6BIc7}%
\end{align}
Separating the purely gravitational terms $\left(  \omega,e\right)  $ from
those containing extra fields $\left(  k,\tilde{k},h,\tilde{h}\right)  $ the
action (\ref{I6BIc7}) can be rewritten as%
\begin{align}
I_{\mathfrak{L}^{\mathfrak{C}_{7}}-BI}^{6D}=\frac{\kappa}{6}\int &  \sigma
_{0}\left[  \epsilon_{abcdef}R^{ab}R^{cd}R^{ef}+\frac{1}{\ell^{6}}%
\epsilon_{abcdef}e^{a}e^{b}e^{c}e^{d}e^{e}e^{f}+\mathcal{\tilde{L}}_{0}\left(
\omega,e,k,\tilde{k},h,\tilde{h}\right)  \right] \nonumber\\
&  +\sigma_{2}\left[  \frac{3}{\ell^{2}}\epsilon_{abcdef}R^{ab}R^{cd}%
e^{e}e^{f}+\mathcal{\tilde{L}}_{2}\left(  \omega,e,k,\tilde{k},h,\tilde
{h}\right)  \right] \nonumber\\
&  +\sigma_{4}\left[  \frac{3}{\ell^{4}}\epsilon_{abcdef}R^{ab}e^{c}e^{d}%
e^{e}e^{f}+\mathcal{\tilde{L}}_{4}\left(  \omega,e,k,\tilde{k},h,\tilde
{h}\right)  \right]  \,. \label{IBIs}%
\end{align}
Omitting the boundary term, each term of the Lovelock series appears in a
different sector of the BI type gravity action. \ The same feature occurs in
the seven-dimensional Chern-Simons case using the $\mathfrak{C}_{7}$ algebra
\cite{CDIMR}. Nevertheless, unlike the odd-dimensional case, the action is not
gauge invariant of the entire gauge group but only under a Lorentz type subgroup.

Let us note that the action (\ref{IBIs}) contains vielbein type fields
$\left(  h,\tilde{h}\right)  $ leading to products of different vielbeins.
This action has an analogous structure to dRGT (de Rham, Gabadadze, Tolley)
massive gravity theories in the vielbein formalism \cite{HR, dRGT14}
suggesting a tri-metric theory in six dimensions, each of them with an EH type
term. A smaller algebra ($\mathfrak{C}_{5}$) would lead only to two kind of
vielbien $(e,h)$ leading to Lagrangian terms analogous to the bi-gravity
formalism \cite{HR1, LI}. It would be interesting to investigate under what
conditions they might be connected to our results. However, important
differences should be pointed out.\textbf{ }Indeed such theories are in the
second order formalism, whilst our description is a first order description.
Additionally, we have\textbf{ }introduced $h$'s only as extra fields and it is
not the original purpose to interpret them as new vielbeins to describe a
n-gravity theory\textbf{. }Besides, since our purpose is to analyze the limits
where those extra fields vanish, we postpone that discussion for future work.

Interestingly diverse gravity actions can be obtained following appropriate
conditions on the $\sigma$'s and applying suitable limits on the extra fields.

\subsubsection{Pure Lovelock gravity action}

The $p=1$ PL action, which trivially corresponds to $EH+\Lambda$, emerges from
the BI type gravity action (\ref{IBIs}) imposing $\sigma_{2}=0,$ $\sigma
_{0}=\sigma_{4}$ and considering a matter-free configuration $\left(
k=\tilde{k}=h=\tilde{h}=0\right)  $%
\begin{equation}
I_{BI\rightarrow PL}^{6D}=\kappa\sigma_{0}\int\frac{1}{2\ell^{4}}%
\epsilon_{abcdef}R^{ab}e^{c}e^{d}e^{e}e^{f}+\frac{1}{6\ell^{6}}\epsilon
_{abcdef}e^{a}e^{b}e^{c}e^{d}e^{e}e^{f}\,,
\end{equation}
where we have omitted the boundary term $\epsilon_{abcdef}R^{ab}R^{cd}R^{ef}$.
This feature is well desired since if BI type gravity theories are the
appropriate theories to describe gravity then such theories should satisfy the
correspondence principle. \ This property will be our principal requirement in
a possible supersymmetric extension of PL theory.

Not only the lowest order PL action can be recovered from the BI type action
but also the maximal one $\left(  p=2\right)  $. \ Indeed, when the
$\sigma_{4}$ constant vanishes and $\sigma_{0}=-\sigma_{2}$, the matter-free
configuration limit $\left(  k=\tilde{k}=h=\tilde{h}=0\right)  $ leads to%
\begin{equation}
I_{BI\rightarrow\max PL}^{6D}=\kappa\sigma_{0}\int\frac{1}{2\ell^{2}}%
\epsilon_{abcdef}R^{ab}R^{cd}e^{e}e^{f}-\frac{1}{6\ell^{6}}\epsilon
_{abcdef}e^{a}e^{b}e^{c}e^{d}e^{e}e^{f}\,.
\end{equation}
As in the CS case, although we obtain the PL action, the dynamical limit
requires a more subtle treatment. \ A detailed discussion about the dynamical
issue\ in odd dimensions has been done in Ref.~\cite{CDIMR}.

\subsubsection{$EH+LL$ gravity action}

Another non-trivial election of the $\sigma$'s leads to an alternative
Lovelock gravity action. The $EH+LL$ action, where $LL$ is an arbitrary
Lanczos-Lovelock term \cite{Deser:2011zk}, can be derived from the BI type
action (\ref{IBIs}) imposing $\sigma_{0}=0,$ $\sigma_{2}=\sigma_{4}$, and
considering a matter-free configuration limit $\left(  k=\tilde{k}=h=\tilde
{h}=0\right)  $:%
\begin{equation}
I_{BI\rightarrow EH+LL}^{6D}=\kappa\sigma_{2}\int\frac{1}{2\ell^{4}}%
\epsilon_{abcdef}R^{ab}e^{c}e^{d}e^{e}e^{f}+\frac{1}{2\ell^{2}}\epsilon
_{abcdef}R^{ab}R^{cd}e^{e}e^{f}\,.
\end{equation}

Thus, the action contains the EH term and the Gauss-Bonnet (GB) term.
Naturally, in six dimensions this is the only possibility to relate the
Einstein-Hilbert term with another higher power in the curvature.
Nevertheless, as we will see in the $2n$-dimensional case, the presence of
higher-curvature terms in the BI type action will allow to equip the EH term
with an arbitrary $p$-order LL term $\left(  p\neq0\right)  $.\ \ 

It is important to point out that the supersymmetric extension of the $EH+LL$
gravity remains unsolved. Although some discussions can be found in
Ref.~\cite{Deser:2011zk}, the explicit form of the supergravity action is
still a mystery. The procedure presented here could be generalized to
supergravity allowing to find a wider class of supergravity theories.

On the other hand, as it was mentioned in Ref.~\cite{CEMZ}, the causality is
violated for quadratic gravity theories when the GB coupling is finite. One
could argue that considering other spin-$2$ fields could fix the causality
violation, however this would lead to restrictive field equations where not
even pp-wave could satisfy. Indeed, the situation is quite different to the
one presented in Ref.~\cite{EHTZ} where no copy of the EH terms appears. In
our present case, the $\mathfrak{C}_{m}$ symmetries imply to copy every
original Lovelock terms avoiding the possibility to find \ Gauss-Bonnet
equations when $h$ or $\tilde{h}$ is identified as the true vielbein. This
does not occur in the PL case where a non-trivial identification of the extra
fields allows to reproduce the appropriate dynamics \cite{CDIMR}.

\subsubsection{Lovelock gravity with generalized cosmological constant}

A six-dimensional gravity action in presence of a generalized cosmological
constant term \cite{SS, AKL} can be recovered in a particular limit. Indeed,
when $\tilde{k}^{ab}=\tilde{h}^{a}=h^{a}=0$, the BI type action reduces to%
\begin{align}
I_{BI\rightarrow EH+LL+\tilde{\Lambda}}^{6D}=  &  \kappa\sigma_{0}\int
\epsilon_{abcdef}\left[  \frac{1}{6l^{6}}e^{a}e^{b}e^{c}e^{d}e^{e}e^{f}%
+\frac{1}{2l^{2}}Dk^{ab}Dk^{cd}e^{e}e^{f}+\frac{1}{2l^{4}}Dk^{ab}e^{c}%
e^{d}e^{e}e^{f}\right] \nonumber\\
&  +\epsilon_{abcdef}\left[  \frac{1}{2l^{2}}R^{ab}R^{cd}e^{e}e^{f}\right]
+\epsilon_{abcdef}\left[  \frac{1}{2l^{4}}R^{ab}e^{c}e^{d}e^{e}e^{f}+\frac
{1}{l^{2}}R^{ab}Dk^{cd}e^{e}e^{f}\right] \nonumber\\
=  &  \kappa\sigma_{0}\int L_{EH}+L_{LL}+L_{\tilde{\Lambda}}\,,
\end{align}
where we have omitted the boundary terms and where we have set $\sigma
_{0}=\sigma_{2}=\sigma_{4}$.

The $L_{\tilde{\Lambda}}$ term includes the cosmological constant term plus
additional pieces depending on the extra field $k^{ab}$.\ This can be seen as
a generalization of the result obtained for the Maxwell symmetry \cite{AKL} to
six dimensions.\ Interestingly, in this limit, the action corresponds to the
six-dimensional BI type constructed out of the curvature two-form for the
$\mathfrak{so}\left(  D-1,2\right)  \oplus\mathfrak{so}\left(  D-1,1\right)  $ algebra.

Let us observe that the same result can be obtained imposing $k^{ab}=0$
instead of $\tilde{k}^{ab}=0$. \ Obviously, the additional terms appearing in
the action would depend in such case on $\tilde{k}^{ab}$. \ The presence of
the extra fields $k^{ab}$ or $\tilde{k}^{ab}$ leads to an alternative
extension of standard gravity allowing the introduction of a generalized
cosmological term. \ At the supersymmetric level, an analogous result has been
presented in Ref.~\cite{CRS} using a supersymmetric extension of the
$\mathfrak{so}\left(  D-1,2\right)  \oplus\mathfrak{so}\left(  D-1,1\right)  $ algebra.

\section{Extension to $D=2n\,\ $gravities}

In this section we present the $2n$-dimensional BI type gravity action based
on the $\mathfrak{C}_{2n+1}$ algebra. \ Additionally, we provide with the
suitable limits on the extra fields and the general conditions on the $\sigma
$'s necessary to recover diverse gravity actions.

Following the same discussion of the previous section, one can see that the
minimal symmetry allowing to separate each Lovelock term in different sectors
corresponds to $\mathfrak{C}_{2n+1}$. From Theorem VII.2 of Ref.~\cite{Sexp},
the non-vanishing components of an invariant tensor necessary to build a
$2n$-dimensional BI type gravity action based on the $\mathfrak{C}_{2n+1}$
curvature two-form are given by%
\begin{equation}
\left\langle J_{a_{1}a_{2},\left(  i_{1}\right)  }\cdots J_{a_{2n-1}%
a_{2n},\left(  i_{n}\right)  }\right\rangle =\frac{2^{n-1}}{n}\alpha
_{2j}\delta_{\left(  i_{1}+\cdots+i_{n}\right)  \operatorname{mod}\left[
n\right]  }^{j}\epsilon_{a_{1}\cdots a_{2n}}\,, \label{ITG}%
\end{equation}
with $i=0,1,\dots,\left[  \frac{2n-1}{2}\right]  $. This invariant tensor
breaks the $\mathfrak{C}_{2n+1}$ symmetry to a Lorentz type subgroup
$\mathfrak{L}^{\mathfrak{C}_{2n+1}}$ generated by $\left\{  J_{ab,\left(
i\right)  }\right\}  $. It is important to clarify that an invariant tensor of
the whole $\mathfrak{C}_{2n+1}$ group would lead to a topological invariant.

Considering the general expression for the curvature two-form (\ref{G2F}) and
the non-vanishing component of the invariant tensor (\ref{ITG}) in the general
expression of a $2n$-dimensional Born-Infeld type gravity action (see eq.
(\ref{BIaction})) we find%
\begin{align}
I_{\mathfrak{L}^{\mathfrak{C}_{2n+1}}\mathfrak{-}BI}^{2n}  &  =\int\sum
_{k=1}^{n}l^{2k-2}\frac{\kappa}{2n}\binom{n}{k}\sigma_{2j}\delta_{\left(
i_{1}+\cdots+i_{n}\right)  \operatorname{mod}\left[  n\right]  }^{j}%
\delta_{p_{1}+q_{1}}^{i_{k+1}}\cdots\delta_{p_{n-k}+q_{n-k}}^{i_{n}%
}\nonumber\\
&  \varepsilon_{a_{1}\cdots a_{2n}}R^{\left(  a_{1}a_{2},i_{1}\right)  }\cdots
R^{\left(  a_{2k-1}a_{2k},i_{k}\right)  }e^{\left(  a_{2k+1},p_{1}\right)
}e^{\left(  a_{2k+2},q_{1}\right)  }\cdots e^{\left(  a_{2n-1},p_{n-k}\right)
}e^{\left(  a_{2n},q_{n-k}\right)  }\,. \label{I2nBI}%
\end{align}
Separating the gravitational terms $\left(  \omega,e\right)  $ from those
containing extra fields $\left(  \omega^{ab,\left(  i\right)  },e^{a,\left(
i\right)  }\text{ for }i\neq0\right)  $, the action (\ref{I2nBI}) can be
rewritten as%
\begin{align}
I_{\mathfrak{L}^{\mathfrak{C}_{2n+1}}\mathfrak{-}BI}^{2n}=\frac{\kappa}%
{2n}\int &  \sigma_{0}\left[  \epsilon_{a_{1}a_{2}\cdots a_{2n}}R^{a_{1}a_{2}%
}\dots R^{a_{2n-1}a_{2n}}+\frac{1}{\ell^{2n}}\epsilon_{a_{1}a_{2}\cdots
a_{2n}}e^{a_{1}}e^{a_{2}}\dots e^{a_{2n}}+\mathcal{\tilde{L}}_{0}\left(
\omega^{ab,\left(  i\right)  },e^{a,\left(  i\right)  }\right)  \right]
\nonumber\\
&  +\sigma_{2}\left[  \frac{n}{\ell^{2}}\epsilon_{a_{1}a_{2}\cdots a_{2n}%
}R^{a_{1}a_{2}}\dots R^{a_{2n-3}a_{2n-2}}e^{a_{2n-1}}e^{a_{2n}}%
+\mathcal{\tilde{L}}_{2}\left(  \omega^{ab,\left(  i\right)  },e^{a,\left(
i\right)  }\right)  \right] \nonumber\\
&  +\cdots\nonumber\\
&  +\sigma_{2n-2}\left[  \frac{n}{\ell^{2n-2}}\epsilon_{a_{1}a_{2}\cdots
a_{2n}}R^{a_{1}a_{2}}e^{a_{3}}\dots e^{a_{2n}}+\mathcal{\tilde{L}}%
_{2n-2}\left(  \omega^{ab,\left(  i\right)  },e^{a,\left(  i\right)  }\right)
\right]  \,.
\end{align}
As in the six-dimensional case, there are diverse configurations allowing to
recover different gravity theories. In the following table we present the
explicit conditions on the $\sigma$'s and the pertinent limit on the extra
fields required to derive diverse gravity actions in $2n$ dimensions:%
\[%
\begin{tabular}
[c]{|c|c|c|}\hline
Gravity actions & $\sigma$'s conditions & Matter-free configuration
limit\\\hline
$p$-order Pure Lovelock & $\sigma_{0}=\left(  \pm1\right)  ^{p+1}%
\sigma_{2n-2p}\neq0,$ $\sigma_{i}=0$ & $\omega^{ab,\left(  i\right)
}=e^{a,\left(  i\right)  }=0$ \ \ for $i\neq0$\\\hline
$EH+p$-order $LL$ & $\sigma_{2n-2}=\sigma_{2n-2\left(  p+1\right)  }\neq0,$
$\sigma_{i}=0$ & $\omega^{ab,\left(  i\right)  }=e^{a,\left(  i\right)  }=0$
\ \ for $i\neq0$\\\hline
$EH+LL+\tilde{\Lambda}$ & $\sigma_{i}=\sigma_{i+1}\neq0$, for $i=0,\dots
,\left[  \frac{2n-3}{2}\right]  $ & $\omega^{ab,\left(  i+1\right)
}=e^{a,\left(  i\right)  }=0$ \ \ for $i\neq0$\\\hline
\end{tabular}
\ \
\]
In particular, $\tilde{\Lambda}$ refers to a generalized cosmological term
which, besides to contain the standard cosmological constant term, includes
additional pieces depending on the extra field $\omega^{ab,\left(  1\right)
}$. \ This generalizes to $2n$ dimensions the results obtained in
Ref.~\cite{SS, AKL}.\ Let us note that, in the six-dimensional case, the extra
field has been identified as $k^{ab}$.

\section{Summary and discussions}

In the present work, we have shown that deforming the BI gravity theory (based
on the $AdS$ curvature) using the abelian semigroup expansion allows to
reproduce diverse Lovelock gravity theories considering appropriate
matter-free configuration limit and imposing pertinent conditions on the
$\sigma$ constants.

The relations among the expanded BI gravity theories and the Lovelock one is
motivated by the recent connection between even-dimensional standard GR and
expanded BI gravity theories \cite{CPRS1, CPRS2} where the properties of the
$S$-expansion were used. \ Diverse results have been obtained, not only in the
BI context, using the Lie Algebra expansion method. \ In fact, it was shown in
Ref.~\cite{EHTZ}, that EH Lagrangian can be derived from an expanded AdS CS
gravity theory. Additionally the proper Einstein dynamics can also be obtained
in an appropriate coupling constant limit using the $S$-expansion procedure
\cite{GRCS}.

The results obtained along this paper not only present some new explicit
relations among BI and Lovelock theories but also could bring valuable
information in order to derive other (super)gravities. Indeed, various
supersymmetric extension of (pure)Lovelock gravity theories\textbf{ }has not
been explored yet and their standard constructions remain to be a highly
difficult task. However, our results suggest that the super (pure)Lovelock
could be related to supergravity theory with expanded Lie algebras. Besides,
the same procedure could be applied to CS supergravity theories.

\section{Acknowledgment}

Research of PKC and EKR is supported by the Newton-Picarte CONICYT Grant No.
DPI20140053. NM was supported by the Chilean FONDECYT Project 3130445. He is
also grateful with the Becas-Chile postdoctoral grant, whose financial support
helped in the last stage of this work. PKC and EKR wish to thank A.
Anabal\'{o}n for his kind hospitality at Departamento de Ciencias of
Universidad Adolfo Iba\~{n}ez. The authors are also grateful with R. Durka for
useful comments.

\end{document}